\documentstyle[prl,aps,epsf,floats]{revtex}

\begin{document}
\draft
\twocolumn[\hsize\textwidth\columnwidth\hsize\csname @twocolumnfalse\endcsname
\title{A class of spin injection-precession ultrafast nanodevices}
\author{V.V. Osipov and A.M. Bratkovsky}
\address{Hewlett-Packard Laboratories, 1501 Page Mill Road, 1L, Palo Alto, CA 94304}
\date{October 8, 2003 }
\maketitle

\begin{abstract}
Spin valve ultrafast spin injection devices are described: an amplifier, a
frequency multiplier, and a square-law detector. Their operation is based on
injection of spin polarized electrons from one ferromagnet to another
through a semiconductor layer and spin precession of the electrons in the
semiconductor layer in a magnetic field induced by a (base) current in an
adjacent nanowire. The base current can control the emitter current between
the magnetic layers with frequencies up to several 100 GHz.
\end{abstract}

\vskip2pc]

\narrowtext

Spintronic devices is a very active area of research \cite
{Wolf}. The devices based on giant- and tunnel magnetoresistance \cite{GMR}
are used in a number of applications. The injection of spin-polarized
electrons into semiconductors is of particular interest because of
relatively large spin-coherence lifetime, and a prospect of using this
phenomenon in spintronic devices and quantum computers \cite{Wolf}. Devices
of that type have been suggested before based on the Rashba spin-orbital
splitting of 2D electrons under a gate electric field \cite{Datta}. The
schemes involving reflection of semiconductor electrons off a ferromagnetic
layer have also been discussed \cite{Ciuti}.

In the present paper we describe a spintronic mechanism of ultrafast
amplification and frequency conversion, which can be realized in
heterostructures comprising a metallic ferromagnetic nanowire surrounded by
a semiconductor (S) and a ferromagnetic (FM) thin shells, Fig.~1(a).
Practical devices may have various layouts, with two examples shown in Fig.
1(b),(c). 
\begin{figure}
\epsfxsize=2.8in \epsffile{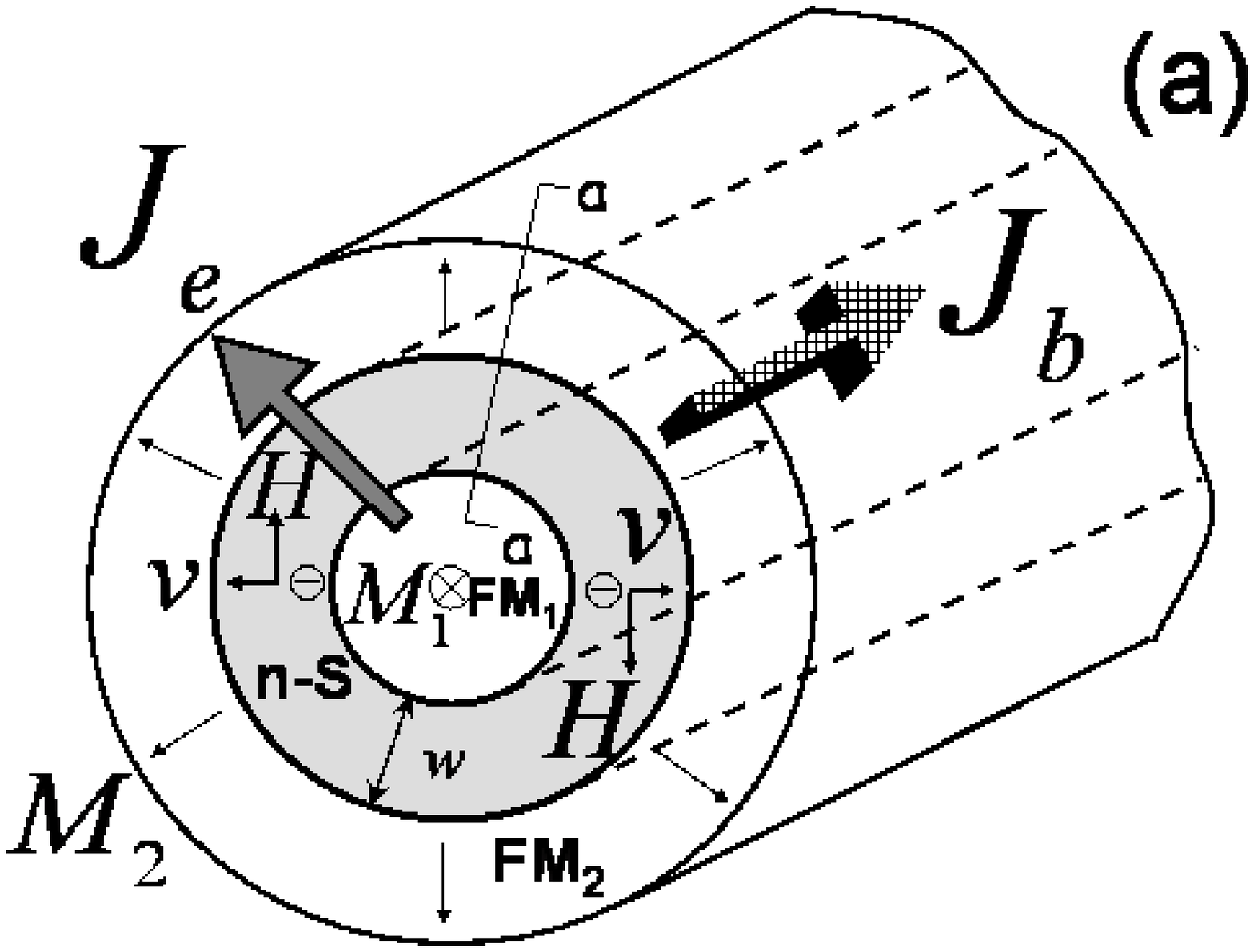}
\vspace{.1in}
\epsfxsize=2.8in \epsffile{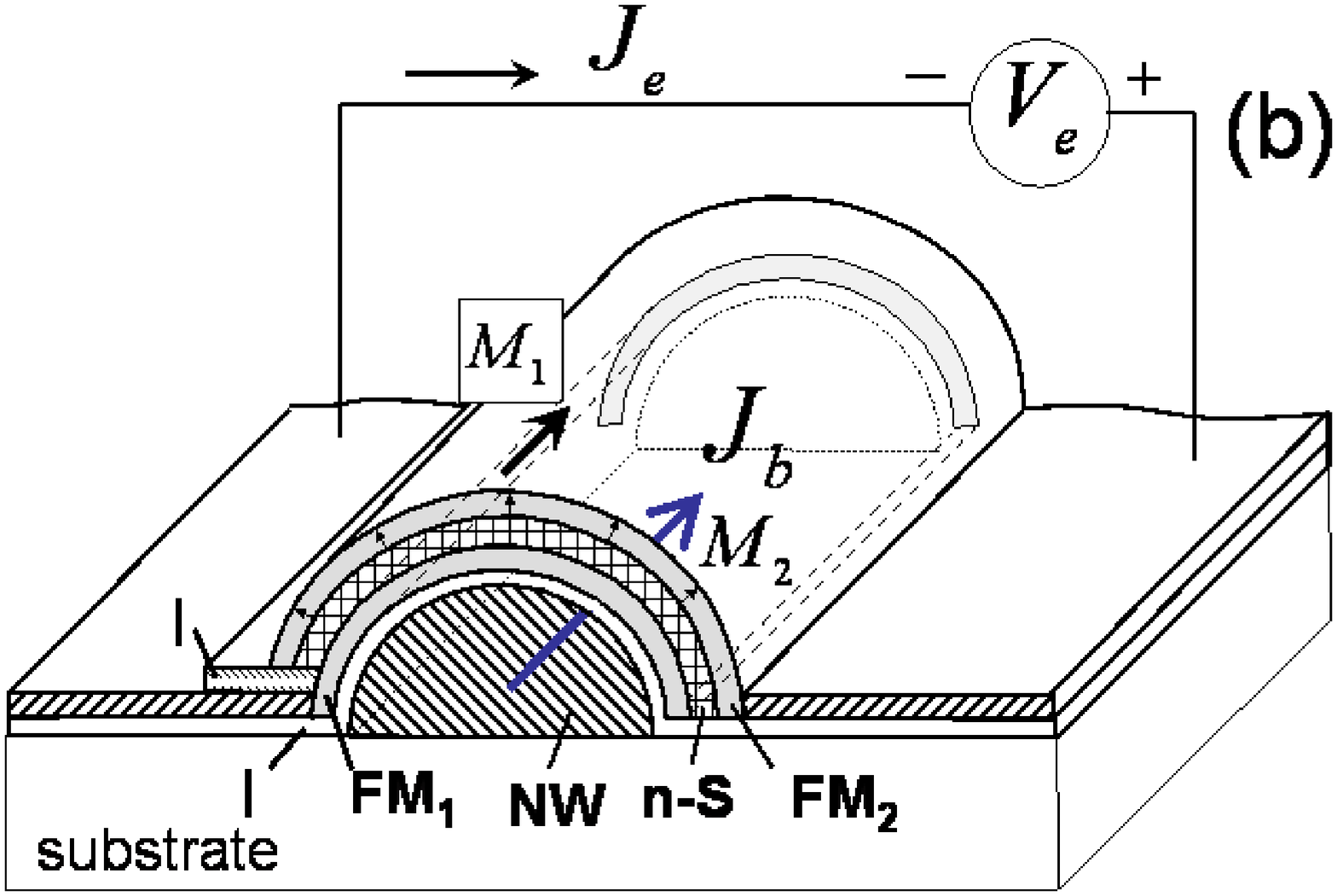}
\vspace{.1in}
\epsfxsize=2.8in \epsffile{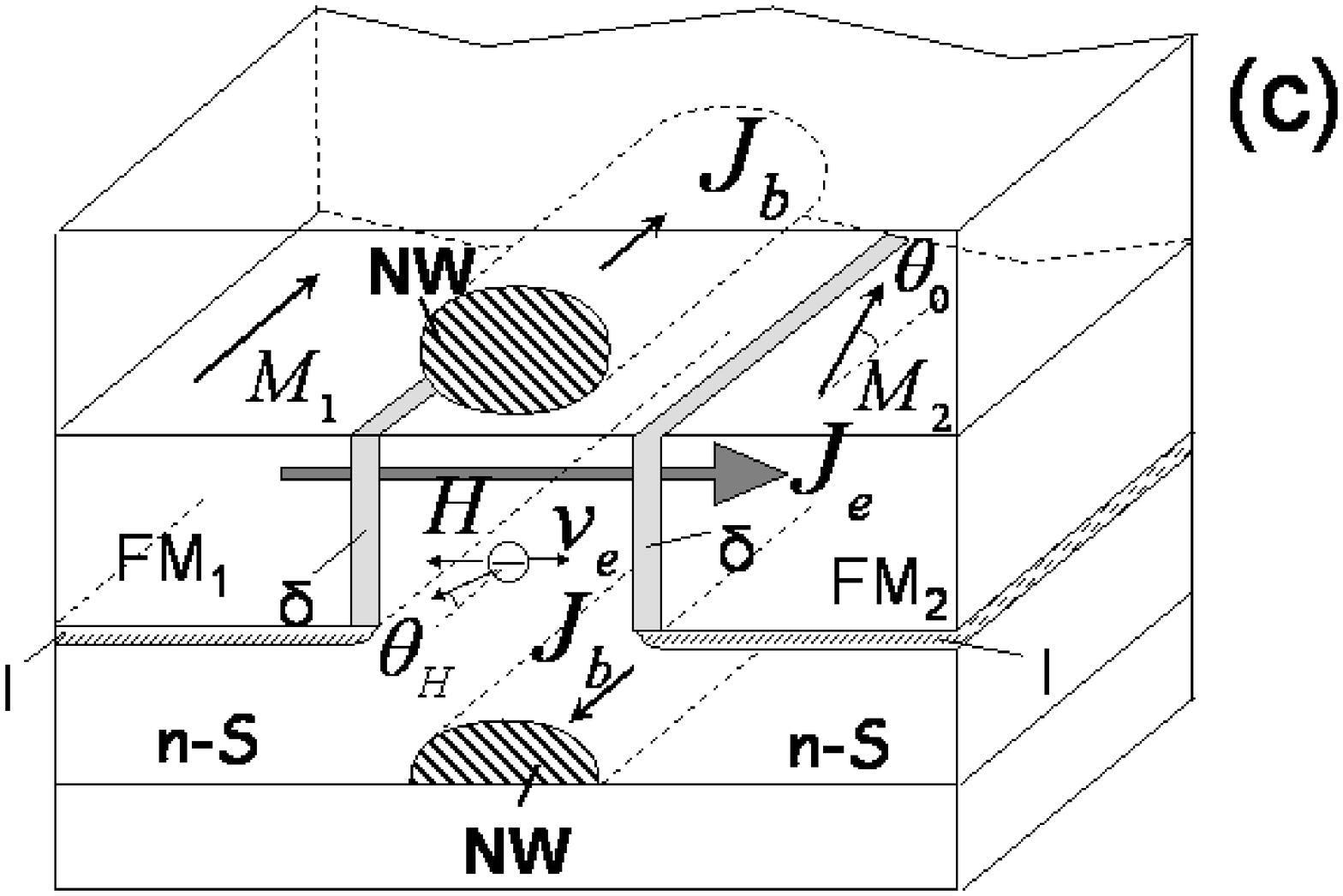}
\vspace{.1in}
\caption{ Schematic of the spin injection-precession devices having
cylindrical (a), semi-cylindrical (b), and planar shape (c). Here FM$_{1}$
and FM$_{2}$ are the ferromagnetic layers; $n-$S the $n-$type semiconductors
layer; $w$ the thickness of the $n-$S layer; $\protect\delta $ the $\protect%
\delta -$doped layers; NW the highly conductive nanowires; $I$ the
insulating layers. The directions of the magnetizations $\vec{M}_{1}$ and $%
\vec{M}_{2}$\ in the FM$_{1}$ and FM$_{2}$ layers, as well as the electron
spin $\protect\sigma $, the magnetic field $H_{b}$, and the angle of spin
rotation $\protect\theta $ in S are also shown. }
\label{fig:fig1}
\end{figure}
Let us consider the principle of operation of the spintronic devices with
the layout shown in Fig.~1(a). We assume that the thickness $w$ of the $n- $%
type semiconductor layer is not extremely small ($w \gtrsim 30$ nm), so that
a tunneling through this layer is negligible. The base voltage $V_{b}$ is
applied between the ends of the nanowire. The base current $J_{b},$ flowing
through the nanowire, induces a cylindrically symmetric magnetic field $%
H_{b}=J_{b}/2\pi \rho $ \ in the S layer, where $\rho $ is the distance from
the center of nanowire. When the emitter voltage $V_{e}$ is applied between
FM layers, the spin polarized electrons are injected from the first layer
(nanowire FM$_{1}$) through the semiconductor layer into the second
(exterior) ferromagnetic shell, FM$_{2}$. We assume that the transit time $%
t_{tr}$ of the electrons through\ the S layer is less than the spin
relaxation time, $\tau _{s}$ (i.e. we consider the case of a spin ballistic
transport). We show below that in this case the emitter current $J_{e}$
between FM$_{1}$ and FM$_{2}$ layers depends on the angle $\theta $ between
the magnetization vectors $\vec M_1$ and $\vec M_2$ in these layers
approximately in the same way as in the tunneling FM-I-FM structures \cite
{Slon,Brat}: 
\begin{equation}
J_{e}=J_{0e}(1+P_{1}P_{2}\cos \theta ).  \label{TC}
\end{equation}
However, in comparison with the tunneling structures, the described
heterostructures have an additional degree of freedom. Spins of injected
electrons will precess in the radially symmetric induced magnetic field $%
H_{b}=J_{b}/(2\pi \rho) $ during the transit of electrons through the
semiconductor layer ($t_{tr}<\tau _{s}$). Therefore, $\theta =\theta
_{0}+\theta _{H}$ in Eq. (\ref{TC}), where $\theta _{0}$ is the angle
between $\vec M_{1}$ and $\vec M_{2}$, and $\theta _{H}$ is the angle of the
spin rotation. The spin precesses with the frequency $\Omega =\gamma
H_{\perp },$ where $H_{\perp }$ is the magnetic field component normal to
the spin and $\gamma $ is the gyromagnetic ratio \cite{LL,SB}. One can see
from Fig. 1(a) that $H_{\perp }=H_{b}$. Thus, the angle of the spin rotation
is equal to $\theta _{H}=\gamma H_{b}t_{tr}=\gamma t_{tr}J_{b}/2\pi \rho
_{S},$ where $\rho _{s}$ is the characteristic radius of the S layer.
According to Eq.~(\ref{TC}), 
\begin{equation}
J_{e}=J_{e0}[1+P_{1}P_{2}\cos (\theta _{0}+k_{j}J_{b})],  \label{JT}
\end{equation}
where $k_{j}=\gamma t_{tr}/2\pi \rho _{S}=\gamma /\omega \rho _{S}$ and $%
\omega =2\pi /t_{tr}$ is the frequency of a variation of the base current, $%
J_{b}=J_{s}\cos(\omega t)$. Equation (\ref{JT}) shows that, when the
magnetization $M_{1}$ is perpendicular to $M_{2}$, $\theta _{0}=\pi /2$, and 
$\theta _{H}\ll \pi ,$ 
\begin{equation}
J_{e}=J_{e0}(1+k_{j}P_{1}P_{2}J_{b}),~G=dJ_{e}/dJ_{b}=J_{e0}k_{j}P_{1}P_{2}.
\label{G}
\end{equation}
Thus, the {\em amplification} of the base current occurs with the gain $G$,
which can be relatively high even for $\omega \gtrsim 100$ GHz. Indeed, $%
\gamma =q/(m_{\ast }c) \approx 2.2(m_{0}/m_{\ast
})10^{5}$ m/A$\cdot$s \cite{LL,SB}, where $m_{0}$ is the free electron mass, $%
m_{\ast }$ the effective mass of electrons in the semiconductor, and $c$ the
velocity of light. Thus, when $\rho _{S}\simeq 30$ nm, $m_{0}/m_{\ast }=14$
(GaAs) and $\omega =100$ GHz, the factor $k_{j}\simeq 10^{3}$ A$^{-1},$ so
that $G>1$ at $J_{e0}>0.1$mA/$(P_{1}P_{2})$.

When $M_{1}$ is collinear with $M_{2}$ ($\theta _{0}=0,\pi )$ and $\theta
_{H}\ll \pi ,$ then, according to Eq.~(\ref{JT}), the emitter current is 
\begin{equation}
J_{e}=J_{e0}(1\pm P_{1}P_{2})\mp \frac{1}{2}
J_{e0}P_{1}P_{2}k_{j}^{2}J_{b}^{2}.\text{\ }  \label{QD}
\end{equation}
Therefore, the time-dependent component of the emitter current $\delta
J_{e}(t)\propto J_{b}^{2}(t)$, and the device operates as a square-law
detector. When $J_{b}(t)=J_{b0}\cos (\omega _{0}t),$ the emitter current has
a component $\delta J_{e}(t)\propto \cos (2\omega _{0}t)$, and the device
works as a {\em frequency multiplier}. When $J_{b}(t)=J_{h}\cos (\omega
_{h}t)+J_{s}\cos (\omega _{s}t),$ the emitter current has the components
proportional to $\cos (\omega _{h}\pm \omega _{s})t$, i.e. the device can
operate as a high-frequency {\em heterodyne detector} with the conversion
coefficient $K=J_{e0}J_{h}P_{1}P_{2}k_{j}^{2}/4 $. For $k_{j}=10^{3}$ A$%
^{-1} $ the value of $K>1$ when $J_{e0}J_{h}>4$(mA)$^{2}/(P_{1}P_{2})$.

Let us now consider the present spintronic devices in greater detail. One
can see from Fig.~1 that the devices in the cross-section $a-a$ are the FM-$%
n$-FM heterostructures, where $n$ marks the $n-$type semiconductor layer.
It is well known that a large potential barrier, so-called depleted Schottky
layer, forms at the metal-semiconductor interface \cite{Sze}. Therefore, the
spin injection current from FM into S in such a FM-$n$-FM structure is
negligibly small when $w\gtrsim 30$ nm. To increase the current, a thin
heavily doped $n^{+}-$semiconductor layer (so-called $\delta $-doped layer)
between the ferromagnet and semiconductor should be used \cite{Jonk,BO}.
This layer screens the interface potential barriers, sharply decreases their
thickness, and increases the tunneling transparency \cite{BO}. This is why
the considered heterostructures have to comprise two $\delta -$doped layers,
between the S and FM layers. Thickness of the $\delta -$doped layers $%
l_{1(2)}$ and the donor concentration $N_{d}^{+}$ there have to satisfy the
following conditions \cite{BO}: $N_{d}^{+}l_{1}^{2}\simeq 2\varepsilon
\varepsilon _{0}(\Delta -\Delta _{0}+rT)/q^{2}$ and $N_{d}^{+}l_{2}^{2}%
\simeq 2\varepsilon \varepsilon _{0}(\Delta -\Delta _{0})/q^{2},$ where $%
\epsilon $ ($\epsilon _{0})$ the dielectric permittivity of semiconductor
(vacuum), $\Delta _{0}=E_{c}-F>0$, $F$ the equilibrium Fermi level, $E_{c%
\text{ }}$ the bottom of a semiconductor conduction band, the parameter $%
r\simeq 2-3,$ and $T$ the temperature (we use the units of $k_{B}=1)$. The
energy diagram of such a FM$-n^{+}-n-n^{+}-$FM structure includes two
potential $\delta -$barriers of the height $(\Delta -\Delta _{0})$ and the
thicknesses $l_{1(2)}$ and a low wide barrier of the height $\Delta _{0}$
and the thickness $w$ in the $n-$semiconductor region (Fig.~2). We assume
that the electrons easily tunnel through the $\delta -$spike barriers due to
a smallness of $l_{1(2)}$. However, only electrons with the energy $E\geq
E_c$ can overcome the low wide barrier $\Delta _{0}$ by way
of a thermionic emission-tunneling \cite{BO}. 
\begin{figure}
\epsfxsize=3.4in \epsffile{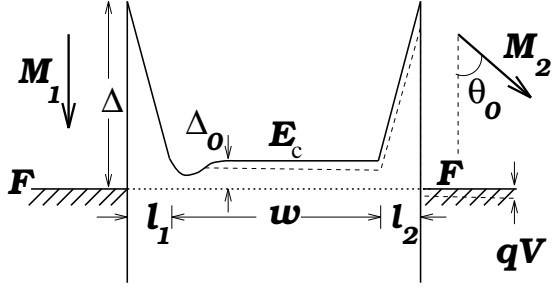}
\caption{Energy diagram of FM$-n^{+}-n-n^{+}-$FM heterostructures in the
cross section $a-a$ in Fig. 1(a). Here $E_{c}$ the bottom of a conduction band
of the semiconductor; $F$ the Fermi level; $\Delta $ ($l_{1(2)})$ the height
(thickness) of the interface potential barrier in the $\protect\delta -$%
doped layers; $\Delta _{0}$ ($w$) the height (width) of the barrier in $n-$%
semiconductor. Broken line shows the bands under relatively small bias
voltage $V$. }
\label{fig:fig2}
\end{figure}
We assume that the electron energy $E$, spin $\sigma $ and the wave vector $%
\vec{k}_{\parallel }$ parallel to the interface are conserved during
tunneling. The current density of electrons with spin $\sigma $ through FM-S
junctions, including the $\delta -$layers, Fig.~2, can be written as \cite
{Duke,Brat,BO} 
\begin{equation}
j_{\sigma }^{1(2)}=\frac{q}{h}\int dE[f(E-F_{\sigma }^{1(2)})-f(E-F_{1(2)})]\int 
\frac{d^{2}k_{\parallel }}{(2\pi )^{2}}T_{k\sigma },  \label{J1-2}
\end{equation}
where $T_{k\sigma }$ is the transmission probability, $\ f(E)=[\exp
(E-F_{1(2)} )/T+1]^{-1}$ the Fermi function, the left (right) Fermi levels are
$F_1=F$ and $F_2=F-qV$, respectively, where $V$ is the bias voltage on
the device, and the integration includes a
summation with respect to a band index. We take into account that the spin
polarized electrons in the semiconductor are out of equilibrium and their
distribution is described by a nonequilibrium Fermi function with the
quasi-Fermi level $F_{\sigma }(x)$. In Eq.~(\ref{J1-2}) $F_{\sigma
}^{1}=F_{\sigma }(0)$ and $F_{\sigma }^{(2)}=F_{\sigma }(w)$. The condition $%
\Delta _{0}=E_{c}-F>0$ means that the semiconductor is nondegenerate, so the
total electron concentration and the concentration of electrons with spin $%
\sigma $ are given by 
\begin{equation}
n=N_{c}\exp \left( -\frac{\Delta _{0}}{T}\right) ,\text{ \ }n_{\sigma }=%
\frac{N_{c}}{2}\exp \left( \frac{F_{\sigma }-E_{c}}{T}\right) ,  \label{Nn}
\end{equation}
where $N_{c}=2M_{c}(2\pi m_{\ast }T)^{3/2}h^{-3}$ is the effective density
of states of the conduction band of the semiconductor, $m_{\ast }$ the
effective mass of electrons in the semiconductor, and $M_{c}$ the number of
the band minima \cite{Sze}. The analytical expressions for $T_{k\sigma }$
can be obtained in an effective mass approximation, $\hbar k_{\sigma
}=m_{\sigma }v_{\sigma }$. For energies of interest, $E\gtrsim F+\Delta
_{0}\mp qV_{1(2)}$, we can approximate the $\delta -$barrier by a triangular
shape and find that, when the voltage drops across the first and the second
junctions $V_{1(2)}$ satisfy the condition $2T\lesssim qV_{1(2)}\lesssim
\Delta _{0}\ll \Delta $, the currents of electrons with spins $\sigma
=\uparrow (\downarrow )$ and $\sigma ^{/}=\pm $ through the junctions of a
unit area are approximately equal to \cite{BO} 
\begin{eqnarray}
J_{1\sigma } &=&J_{01}d_{\sigma }\exp (qV_{1}/T),\text{ }J_{2\sigma ^{\prime
}}=2J_{02}d_{\sigma ^{\prime }}n_{\sigma ^{\prime }}(w)/n,  \label{1} \\
J_{01(2)} &=&-\alpha _{1(2)}^{0}qv_{T}n\exp \left( -\eta \kappa
_{0}^{1(2)}l_{1(2)}\right) ,  \label{ds}
\end{eqnarray}
where the spins $\sigma =\uparrow $ and $\sigma ^{\prime }=+$ are parallel
to the magnetizations $\vec{M}_{1}$ and $\vec{M}_{2}$, respectively, 
\begin{eqnarray}
\kappa
_{0}^{1(2)}\equiv 1/l_{0}^{1(2)}&=&(2m_{\ast }/\hbar ^{2})^{1/2}(\Delta
-\Delta _{0}\pm qV_{1(2)})^{1/2},\nonumber \\
\alpha _{1(2)}^{0}&=&1.2\left( \kappa
_{0}^{1(2)}l_{1(2)}\right) ^{1/3},\nonumber \\ 
d_{\sigma (\sigma ^{\prime
})}&=&v_{T}v_{\sigma (\sigma ^{\prime })}/\left( v_{t1(t2)}^{2}+v_{\sigma
(\sigma ^{\prime })}^{2}\right), \nonumber \\ 
v_{\sigma (\sigma ^{\prime })}&=&v_{\sigma
(\sigma ^{\prime })}(\Delta _{0}\mp qV_{1(2)}),\nonumber\\ 
v_{t1(t2)}&=&\sqrt{2(\Delta -\Delta _{0}\pm qV_{1(2)})/m_{\ast
}}\nonumber
\end{eqnarray}
and 
$$v_{T}\equiv \sqrt{3T/m_{\ast }}.$$

In the case of a spin ballistic transport ($t_{tr}<\tau _{s}$), the spin of
injected electrons is conserved in the semiconductor layer, $\sigma^\prime
=\sigma $, where the spins $\sigma $ is determined by the direction $\vec{M}%
_{1}.$ Therefore, the angle between the spin $\sigma $ and the
magnetization\ $\vec{M}_{2}$ is $\theta =\theta _{0}+\theta _{H}$, where $%
\theta _{0}$ is the angle between $\vec{M}_{1}$ and $\vec{M}_{2}\ $and $%
\theta _{H}$ the angle of spin precession in the magnetic field $%
H_{b}=J_{b}/2\pi \rho_S $. Probabilities of an electron spin $\sigma $ to have
a projection onto the axes $\pm \vec{M}_{2}$ are $\cos^2(\theta /2)$ and $%
\sin ^{2}(\theta /2)$, respectively \cite{LL}. Therefore, using Eqs. (\ref{1}%
), the spin current through the second (right) junction can be written as 
\begin{equation}
J_{2\uparrow (\downarrow )}=J_{02}\frac{2n_{\uparrow (\downarrow )}(w)}{n}%
\left( d_{+(-)}\cos ^{2}\frac{\theta }{2}+d_{-(+)}\sin ^{2}\frac{\theta }{2}%
\right).  \label{down}
\end{equation}
Considering the total current density $J=J_{1\uparrow }+J_{1\downarrow
}=J_{2\uparrow }+J_{2\downarrow },$ it follows from Eqs.~(\ref{1}) and (\ref
{down}) that 
\begin{eqnarray}
J_{1\uparrow } &=&(J/2)(1+P_{1}),  \label{J11} \\
J_{2\uparrow } &=&\frac{J}{2}\frac{(1+2\delta n_{\uparrow
}(w)/n)(1+P_{2}\cos \theta )}{1+(2\delta n_{\uparrow }(w)/n)P_{2}\cos\theta }%
.  \label{J22}
\end{eqnarray}
Here $P_{1(2)}$ are the spin factors 
\begin{equation}
P_{1(2)}=\frac{d_{\uparrow (+)}-d_{\downarrow (-)}}{d_{\uparrow
(+)}+d_{\downarrow (-)}}=\frac{(v_{\uparrow (+)}-v_{\downarrow
(-)})(v_{t1(2)}^{2}-v_{\uparrow +}v_{\downarrow -})}{(v_{\uparrow
(+)}+v_{\downarrow (-)})(v_{t1(2)}^{2}+v_{\uparrow +}v_{\downarrow -})},
\label{P12}
\end{equation}
which coincide with spin polarization of a current in tunneling FM-I-FM
structures \cite{Brat}.

The spatial distribution of spin polarized electrons is determined by the
usual equation 
$
dJ_{\sigma }/dx=q\delta n_{\sigma }/\tau _{s},  
$
where $\delta n_{\sigma }=n_{\sigma }-n/2$
\cite{Sze,Flat}. Integrating this equation over  
$x$ along the semiconductor layer of thickness $w$ we obtain $J_{1\uparrow
}-J_{1\uparrow }=q\tau _{s}^{-1}\int \delta n_{\sigma }dx<qnw/\tau _{s}$.
Therefore, when $J_{1\uparrow }\gg J_{s}=qnw/\tau _{s}$ and $J_{1\uparrow
}\gg J_{s}$, Eqs.~(\ref{J11}),(\ref{J22}) yield 
\begin{eqnarray}
&&2\delta n_{\uparrow }(w)/n =(P_{1}-P_{2}\cos \theta )/(1-P_{1}P_{2}\cos
\theta )  \label{n00} \\
&&J =J_{02}(d_{+}+d_{-})\left( 1-P_{2}^{2}\cos ^{2}\theta \right)
(1-P_{1}P_{2}\cos \theta )^{-1}\text{.}  \label{Jtet}
\end{eqnarray}
One can see that Eq.~(\ref{Jtet})
gives the same qualitative behavior as Eqs.~(3),(4),
so indeed the effects described above can be realized in the
heterostructures shown in Fig.~1.

We note that both $\kappa ^{1(2)}$ and $P_{1(2)}$ are the functions of the
bias $\Delta _{0}$ and $V_{1(2)}$. Therefore, by adjusting $\Delta _{0}$ and 
$V_{1(2)}$ one can maximize a polarization of the injected current \cite{BO}. 
This may
be achieved when the bottom of conduction band in a semiconductor, $
E_{c}$, is close to a peak in the density of
(minority) electron states in the elemental ferromagnet like Fe, Co, Ni (cf. 
\cite{BO}, for example, in Ni and Fe $\Delta _{\downarrow }\simeq 0.1$ eV 
\cite{Mor}).


\end{document}